\documentclass[preprint,superscriptaddress]{revtex4-1}

\usepackage{bm}
\usepackage{braket}
\usepackage{amsmath}
\usepackage[dvips]{graphicx}
\begin{document}


\title{Current-induced Orbital and Spin Magnetizations in Crystals with Helical Structure}



\author{Taiki Yoda}
\affiliation{Department of Physics, Tokyo Institute of Technology, Tokyo 152-8551, Japan}
\author{Takehito Yokoyama}
\affiliation{Department of Physics, Tokyo Institute of Technology, Tokyo 152-8551, Japan}
\author{Shuichi Murakami}
\email{murakami@stat.phys.titech.ac.jp}
\affiliation{Department of Physics, Tokyo Institute of Technology, Tokyo 152-8551, Japan}
\affiliation{
 TIES, Tokyo Institute of Technology, 2-12-1 Ookayama, Meguro-ku, Tokyo 152-8551, Japan
}

\begin{abstract}
We theoretically show that in a crystal with a helical lattice structure, orbital and spin magnetizations along a helical axis are induced by an electric current along the helical axis.
We propose a simple tight-binding model for calculations,  and the results
can be generalized to any helical crystals. The induced magnetizations are opposite for right-handed and left-handed helices. 
The current-induced spin magnetization along the helical axis comes from a radial spin texture on the Fermi surface. 
This is in sharp contrast to Rashba systems where the induced spin magnetization is perpendicular to the applied current.
\end{abstract}

\date{\today}


\maketitle

Recent discovery of novel physics due to an interplay between electricity and magnetism such as multiferroics~\cite{Spaldin15072005,0022-3727-38-8-R01,eerenstein2006multiferroic,cheong2007multiferroics}, skyrmion~\cite{rossler2006spontaneous,Muhlbauer13022009,fert2013skyrmions,nagaosa2013topological}, and current-induced magnetization reversal~\cite{mangin2006current,PhysRevB.78.212405,chernyshov2009evidence,miron2010current}, 
makes it possible to control a magnetization with electric field.
However, there is a well-known classical example; an electric current flowing through a solenoid induces a magnetization. 
In this paper, we theoretically propose a condensed-matter analogue of a solenoid.
We consider a three-dimensional crystal with a helical lattice structure.
For crystals with helical crystal structure such as Se or Te, one can define handedness similar to a solenoid. Keeping these crystals in mind, we propose a simple tight-binding model with handedness, which can grasp the essence of the physics of helical crystals. 
Then we show that orbital and spin magnetization along a helical axis is induced  by an electric current along the helical axis. 
The induced magnetizations are opposite for right-handed and left-handed helices. 
In contrast to Rashba system, the spin texture on the Fermi surface is radial, 
and the  current-induced spin magnetization along the helical axis in helical crystals 
is attributed to 
this radial spin texture. Our results
can be generalized to any crystals without mirror and inversion symmetries and they would pave the way to spintronics application of helical crystals.

\begin{figure}[b]
\includegraphics[width=8.6cm]{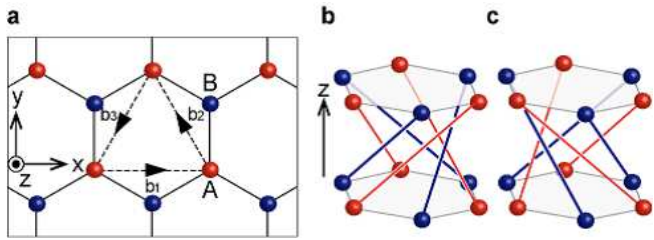}
\caption{{\bf Helical lattice structure of the present model}. {\bf a}, One layer of the model, forming a honeycomb lattice. Dashed arrows denote vectors ${\bf b}_{1}, {\bf b}_{2},$ and ${\bf b}_{3}$. {\bf b}, Hopping texture in the right-handed helix. {\bf c}, Hopping texture in the left-handed helix. Red (blue) balls are hoppings between A (B) sites.}
\label{fig_honeycomb}
\end{figure}

First, we consider an orbital magnetization in a helical crystal. Here, we introduce a three-dimensional tight-binding model with a right- or left-handed helical structures. The lattice structure of this model is composed of an infinite stack of honeycomb lattice layers with one orbital per site. We consider the honeycomb lattice as shown in Fig.\ref{fig_honeycomb}{\bf a}, 
with ${\bf b}_{1} =a\hat{\bf x}$, ${\bf b}_{2} =a/2(-\hat{{\bf x}} + \sqrt{3}\hat{{\bf y}})$, and ${\bf b}_{3} = -{\bf b}_{1}-{\bf b}_{2}$,  
where $a$ is a constant. The 
layers are stacked along the $z$-direction with the primitive lattice vector ${\bf a}_{3} = c\hat{{\bf z}}$, where $c$ is the interlayer spacing.
The Hamiltonian is
\begin{equation}
H = t_{1}\sum_{\langle ij \rangle}c^{\dagger}_{i}c_{j} + t_{2}\sum_{[ ij ]}c^{\dagger}_{i}c_{j} + \Delta\sum_{i}\xi_{i}c^{\dagger}_{i}c_{i},
\label{spinless_Hamiltonian}
\end{equation}
where $t_{1}$, $t_{2}$ and $\Delta$ are real and we set $t_1>0$ for simplicity. The first term is a nearest-neighbor hopping term within the $xy$ plane. The second term represents ``helical" hoppings between the same sublattice in the neighboring layers. This term is different between a right-handed and a left-handed helices as 
shown in Fig.\ref{fig_honeycomb}{\bf b} and Fig.\ref{fig_honeycomb}{\bf c}. 
In the right-handed helix (Fig.\ref{fig_honeycomb}{\bf b}), the direction of hoppings between A sites are $\pm ({\bf b}_{i} + c\hat{{\bf z}} )$, and that 
between B sites are $\pm ( -{\bf b}_{i} + c\hat{{\bf z}} )$. 
Similarly in the left-handed helix (Fig.\ref{fig_honeycomb}{\bf c}) the direction of hoppings between A sites are  $\pm ( {-\bf b}_{i} + c\hat{{\bf z}} )$, and that between B sites are $\pm ( {\bf b}_{i} + c\hat{{\bf z}} )$. This term breaks inversion and mirror symmetries. The third term is a staggered on-site potential, where $\xi_{i}$ is $+1$ and $-1$ for the sites in A and B sublattices, respectively. The space group of our model is {\it P}321, 
with $C_{2y}$ and $C_{3z}$ symmetries.
Only when $\Delta = 0$, the space 
group becomes {\it P}622, having 
additional $C_{2x}$ and $C_{6z}$ symmetries.

\begin{figure}[t]
\includegraphics[width=8.6cm]{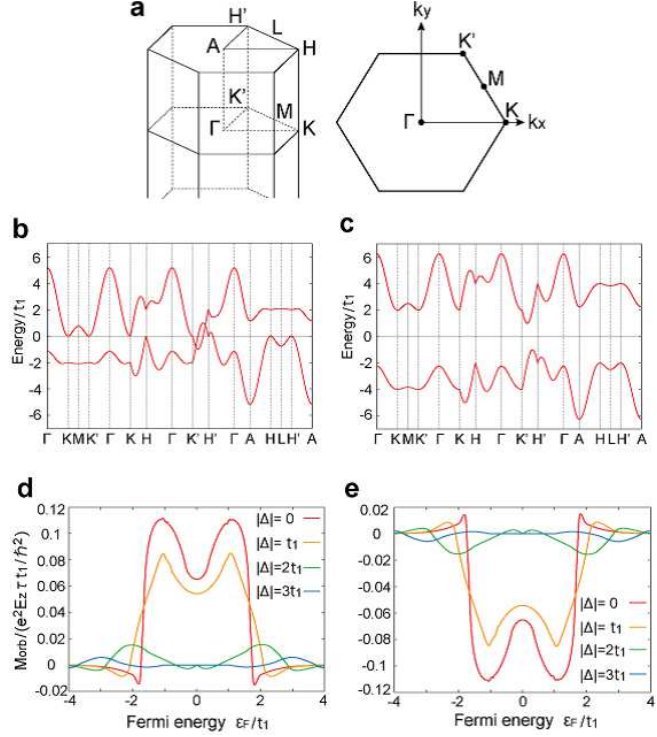}
\caption{{\bf Current-induced orbital magnetization}. {\bf a}, Brillouin zone of our model with high-symmetry points and that on $k_{z} = 0$. {\bf b}, {\bf c}, Energy bands for the Hamiltonian (\ref{spinless_Hamiltonian}) with $t_{2} = t_1/3$, and {\bf b} $\Delta = t_1$; {\bf c} $\Delta = 3t_1$. The energy bands are indicated within $0 \le k_{z} \le \pi/c$. {\bf d}, {\bf e}, Orbital magnetization for several values of $\Delta$ in {\bf d} the right-handed helix  and {\bf e} the left-handed helix. 
We set the parameters $t_{2} = t_1/3$.}
\label{fig_Brillouin_zone_and_energy_band}
\end{figure}

 We note that the Hamiltonian for the right-handed (left-handed) helix
becomes the Haldane model on a honeycomb lattice~\cite{PhysRevLett.61.2015} by replacing +$k_{z}c$ ($-k_z c$) with a Aharonov-Bohm phase $\phi$ in the second-neighbor hopping in the Haldane model. We also note that the time-reversal symmetry is present in our model, while it is absent in the Haldane model for $\phi \neq 0$.
Hence the band structure of our model is easily obtained from that of the Haldane model~\cite{PhysRevLett.61.2015}.
The Brillouin zone (BZ) and band structure of our model are shown in Figs. \ref{fig_Brillouin_zone_and_energy_band}{\bf a} and, {\bf b} and {\bf c}, respectively. In the BZ, the K (K') point is defined by ${\bf k} \cdot {\bf b}_{i} = -2\pi/3$ ($2\pi/3$) on the $k_{z} = 0$ plane, and the H (H') point is similarly defined on the $k_{z} = \pi/c$ plane.

%

Using this model, we calculate the orbital magnetization induced by an electric field along the helical axis. 
In the limit of zero temperature $T \rightarrow 0$, it is calculated from the formula~\cite{PhysRevLett.95.137204,PhysRevLett.95.137205,PhysRevB.74.024408,PhysRevLett.97.026603,
PhysRevLett.99.197202,RevModPhys.82.1959,0953-8984-22-12-123201}
\begin{eqnarray}
	{\bf M}_{\text{orb}} &=& 2\sum_{n} \int_{\text{BZ}} \frac{d^{3}k}	{(2\pi)^{3}}f_{n{\bf k}}\Bigl[{\bf m}_{n{\bf k}} + \frac{e}{\hbar}(\varepsilon_{F} - \varepsilon_{n{\bf k}}){\bf \Omega}_{n{\bf k}}\Bigr],
	\label{orbital_magnetization_1}
\end{eqnarray}
where the integral is performed over the BZ,  $n$ denotes the band index, $f_{n{\bf k}}$ is the distribution function for the eigenenergy $\varepsilon_{n{\bf k}}$, $\varepsilon_{F}$ is the Fermi energy, and the prefactor 2 comes from the spin degeneracy. The orbital magnetic moment of the Bloch electrons is defined by~\cite{PhysRevB.53.7010,PhysRevB.59.14915}
${\bf m}_{n{\bf k}} = -i(e/2\hbar)\Braket{\partial_{{\bf k}}u_{n{\bf k}} | \times [H_{{\bf k}} - \varepsilon_{n{\bf k}}] | \partial_{{\bf k}}u_{n{\bf k}} }$
and the Berry curvature is defined by
${\bf \Omega}_{n{\bf k}} = i\Braket{\partial_{{\bf k}} u_{n{\bf k}} | \times |\partial_{{\bf k}} u_{n{\bf k}}}$,
where $\ket{u_{n{\bf k}}}$ is the periodic part of the Bloch function. Because our model lacks inversion symmetry, Both
${\bf m}_{n{\bf k}}$ and ${\bf \Omega}_{n{\bf k}}$
are allowed to have nonzero values for any ${\bf k}$. However, in equilibrium, due to the time-reversal symmetry ${\bf M}_{\mathrm{orb}}$ is zero because in Eq. (\ref{orbital_magnetization_1}) the contribution from ${\bf k}$ and that from $-{\bf k}$ cancel each other. To induce the orbital magnetization, we apply an electric field along the $z$ axis. For a metal it induces a charge current accompanied by nonequilibrium electron distribution, and orbital magnetization is expected to arise. Within the Boltzmann approximation, the applied electric field $E_{z}$ changes $f_{n{\bf k}}$ into
\begin{equation}
f_{n{\bf k}} = f^{0}_{n{\bf k}} + eE_{z}\tau v_{n,z} \left. \frac{d f}{d \varepsilon} \right|_{\varepsilon = \varepsilon_{n{\bf k}}}
\label{Boltzmann_equation}
\end{equation}
where
$f^{0}_{n{\bf k}} = f(\varepsilon_{n{\bf k}})$ is the Fermi distribution function in equilibrium, $\tau$ is the relaxation time assumed to be constant, and 
$v_{n,z} = (1/\hbar) \partial \varepsilon_{n{\bf k}}/\partial k_{z} $
is the velocity in the $z$ direction. Substituting $f_{n{\bf k}}$ into 
Eq. (\ref{orbital_magnetization_1}), we obtain
the current-induced orbital magnetization ${\bf M}_{\mathrm{orb}}$, which is 
along the $z$-axis by symmetry. 
Because of time-reversal symmetry, the $f^{0}_{n{\bf k}}$ term in Eq. (\ref{Boltzmann_equation}) does not contribute to ${\bf M}_{\mathrm{orb}}$. 
In the $T \rightarrow 0$ limit, $\partial f/\partial \varepsilon$ has a sharp peak at $\varepsilon_{F}$. Therefore, for a band insulator, the induced orbital magnetization ${\bf M}_{\mathrm{orb}}$ is zero. This is in contrast to multiferroics. 
Furthermore, the Berry-curvature term (second term in Eq. (\ref{orbital_magnetization_1})) always vanishes at $T \rightarrow 0$.
%

Figure \ref{fig_Brillouin_zone_and_energy_band} {\bf d}, {\bf e} shows numerical results of Eq. (\ref{orbital_magnetization_1})
for several values of $\Delta$. We set the parameters as $t_{2} = t_1/3$.
The band structure is easily obtained from the similarlity to the Haldane model~\cite{PhysRevLett.61.2015}.
For $|\Delta|>2t_1$, the bands have a finite gap. For $\sqrt{3}t_1 < |\Delta| < 2t_1$, the two bands overlap and the system is in the semimetal phase. For $|\Delta| \leq \sqrt{3}t_1$, the two bands cross each other only on K-H line and K'-H' line at $\Delta = -\sqrt{3}t_1\sin k_{z}c$ and $\Delta = \sqrt{3}t_1\sin k_{z}c$, forming Dirac cones at energies $\pm t_{1}\sqrt{1 - (\Delta^{2}/3t_{1}^{2})}$, 
respectively.
 As is expected, when the Fermi energy lies in the energy gap ($\Delta = 3t_1$, $-t_1 < \varepsilon_{F}<t_1$  in Fig.\ref{fig_Brillouin_zone_and_energy_band}{\bf d}, {\bf e}), the orbital magnetization is zero for $T \rightarrow 0$. For metals ($\Delta = 0$ and $\Delta = t_1$), the current-induced orbital magnetization appears. In this case, the induced magnetization is largely enhanced around $\varepsilon_{F} \simeq t_1$. This is attributed to an enhanced
orbital magnetic moment ${\bf m}_{n{\bf k}}$ near the Dirac points appearing on the K-H or K'-H' lines. 
The orbital magnetization in the left-handed helix (Fig.\ref{fig_Brillouin_zone_and_energy_band}{\bf e}) is exactly opposite to that in the right-handed one (Fig.\ref{fig_Brillouin_zone_and_energy_band}{\bf d}). 
This dependence on handedness  is similar to the solenoid, where an electric current generates a magnetic field.
Thus, based on the semiclassical theory, the orbital magnetization induced by the current is attributed to 
 a helical motion of a wavepacket.
Here, we note that although most of the magnetoelectric effect involve the spin-orbit interaction, the current-induced orbital magnetization predicted here appears even without the spin-orbit interaction. 


Next, we consider current-induced spin polarization. To this end, we introduce the spin-orbit interaction into a tight-binding model with the helical structure:
\begin{eqnarray}
H_{\text{so}} = t_{1 }\sum_{\langle ij \rangle}c^{\dagger}_{i}c_{j} + (i \sqrt3 \lambda / a) \sum_{\langle ij \rangle} c^{\dagger}_{i} (s^{x} d_{ij}^{x}  + s^{y}d_{ij}^{y}) c_{j}  \nonumber
\\
+ (i \lambda_{xy}/a) \sum_{[ ij ]} c^{\dagger}_{i} (s^{x} d_{ij}^{x}  + s^{y}d_{ij}^{y}) c_{j} \nonumber 
\\
+ (i \lambda_{z}/c) \sum_{[ ij ]} c^{\dagger}_{i} s^{z} d_{ij}^{z} c_{j} + \Delta \sum_{i}\xi_{i}c^{\dagger}_{i}c_{i}.
\label{spin_Hamiltonian}
\end{eqnarray}
The first term is a spin-independent nearest-neighbor hopping term within the $xy$ plane and the fifth term is a staggered on-site potential term. The other three terms represent spin-orbit interactions ($s$ are the Pauli matrices in spin space). The second term is a spin-dependent nearest-neighbor hopping term within the $xy$ plane. The third and forth term involve spin-dependent helical hoppings between the neighboring layers as shown in Figs.\ref{fig_honeycomb}{\bf b} and {\bf c}, where $d_{ij}^{k}$ is the $k$-component of a vector pointing from site $j$ to site $i$. The space group of this model is the same as that of the model (\ref{spinless_Hamiltonian}).
%

\begin{figure}
\includegraphics[width=6.5cm]{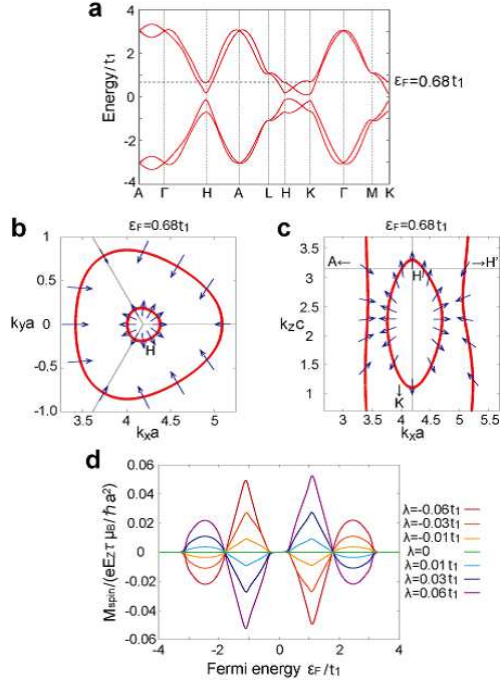}
\caption{{\bf Current-induced spin magnetization}. {\bf a}, Energy band for the Hamiltonian with $\lambda = -0.06t_1, \lambda_{xy} = \lambda_{z} = 0.05t_1$, and $\Delta = 0.4t_1$. Dashed line indicates $\varepsilon_F = 0.68t_1$. {\bf b}, {\bf c}, Fermi surface and spin texture. The parameters are the same as {\bf a}. Arrows represent the spin projected onto each plane. The spin on the inner Fermi surface in {\bf b} is drawn with three times larger scale.
{\bf d}, Spin magnetization for several values of $\lambda$. The parameters are $\lambda_{xy} = 0.05t_1$, $\lambda_{z} = 0.05t_1$, and $\Delta = 0.4t_1$. We also set the temperature $T = 0.03t_1/k_{\text{B}}$.}
\label{fig_Energy_band_and_spin_texture}
\end{figure}

Figure \ref{fig_Energy_band_and_spin_texture}{\bf a} shows the band structure of the Hamiltonian (\ref{spin_Hamiltonian}) along some high symmetry lines. Figure \ref{fig_Energy_band_and_spin_texture}{\bf b} shows the spin texture 
projected onto the $xy$ plane, 
on the Fermi surface around the H point at $\varepsilon_{F} = 0.68t_1$. 
In fact, the spin around the H point has not only the $xy$-component but also the $z$-component.
Because of the spin-orbit interaction, the two spin-split Fermi surfaces appear, having the opposite spin orientations. Remarkably, unlike Rashba systems, the spin is oriented radially, and rotates once around the H point. 
Figure \ref{fig_Energy_band_and_spin_texture}{\bf c} shows the spin texture between the K and H points at $\varepsilon_{F} = 0.68t_1$. In this case, 
one of the spin-split Fermi surface is open. Nevertheless, the inner Fermi surface have a radial spin texture. This spin texture results from crystal symmetries. Namely, 
the K-H lines are three-fold rotation axis and hence the spin on these lines are parallel to these K-H lines.
Furthermore, 
the absence of mirror symmetries is crucial for the radial spin textures; if a mirror plane including the $xy$ plane or the $z$ axis were present, the spins should be perpendicular to the mirror plane, and 
a radial spin texture would not appear. 
For example, Te and Se have helical crystal structure, and as expected, they have been predicted to show a radial spin texture~\cite{hirayama2014weyl}.
%

We now calculate current-induced spin magnetization in the present system. 
It is known that in Rashba systems the current-induced magnetization is perpendicular 
to the current, because of the tangential spin texture in the spin-split Fermi surfaces \cite{edelstein1990spin,PhysRevB.67.033104,PhysRevLett.93.176601,sih2005spatial}. Similarly, a spin magnetization is induced by an electric current in our model, whereas the magnetization is parallel to the current when the current is along the helical axis by symmetry, thanks to 
the radial spin texture. 
Within the Boltzmann approximation as explained in the Methods section, 
the $z$-component of ${\bf M}_{\text{spin}}$ for several parameters
is numerically calculated 
as shown in Figure \ref{fig_Energy_band_and_spin_texture}{\bf d}. 
We fix $\lambda_{xy} = 0.05t_1$, $\lambda_{z} = 0.05t_1$, and $\Delta = 0.4t_1$. As is expected, for a metal, the spin magnetization parallel to the current is induced by the current. Similar to the orbital magnetization, the spin magnetization in the left-handed helix and that in the right-handed helix are opposite. 

We have shown that in systems with helical structure, the orbital and spin magnetizations are induced by an electric current along the helical axis, using a simple tight-binding models. When an electric field is applied along the helical axis, the orbital magnetization is induced by the orbital magnetic moment on the Fermi surface, as a consequence of broken inversion symmetry. 
Similarly to the Berry curvature, the orbital magnetic moment is enhanced near band crossings. Therefore, when the Fermi energy lies near the band crossing, the orbital magnetization is enhanced as well. 
We have also shown that, in the helical crystal with spin-orbit interaction, a current induces the spin magnetization along the helical axis 
due to the radial spin texture. The absence of mirror symmetry allows to have the radial spin texture, which is completely different from Rashba systems with a tangential spin texture.

We have presented a toy model for a helical structure. 
From a symmetry viewpoint, the present effect of 
longitudinal current-induced magnetization 
appears for chiral crystals without inversion and mirror symmetries. 
Real helical crystals, such as Se and Te, also lack inversion and mirror symmetries. From symmetry consideration, we expect that orbital and spin magnetizations along the helical axis can be induced by an electric current in these crystals by doping carriers. 
Magnitudes of the induced magnetizations can be estimated from our results. 
The induced orbital magnetization scales with $e^2E_z\tau t_1/\hbar^2$, which is about 
150 gauss for $E_z=10^4$V/m, $\tau=10^{-12}$s and $t_1=3$eV. In Fig.~\ref{fig_Brillouin_zone_and_energy_band}, the maximum value is $0.1$ times the above 
scale, and this factor could be enhanced for more ``helical'' crystalline structure.
On the other hand, the induced spin magnetization scales wtih $eE_z\tau\mu_B/\hbar a^2$, which is about 
7 gauss for $E_z=10^4$V/m, $\tau=10^{-12}$s and $a=0.5$nm. It is multiplied by a numerical factor in Fig.~\ref{fig_Energy_band_and_spin_texture}, approximately given by the 
ratio $\lambda/t_1$. Thus, the size of the spin-orbit coupling limits the 
size of the current-induced spin magnetization. On the other hand, the orbital magnetization does not require spin-orbit coupling, 
and it can be enhanced by appropriate choice of materials with helical crystal structure. Te and Se consist of weakly coupled helices, and therefore they may have large current-induced orbital magnetization.

Our results also provide a new building block of spintronics and will pave the way to spintronics application of helical crystals. 
For example, consider a ferromagnet on a helical crystal. By injecting current into the helical crystal, the induced magnetization exerts a torque on the ferromagnet, thus leading to current-induced magnetization reversal. 

\vspace{2mm}

{\small
{\bf Methods:}\ 
{\bf Details of the calculation of current-induced spin magnetization}--
Within the Boltzmann approximation, the electric field $E_{z}$ induces the spin magnetization as
\begin{eqnarray} 
{\bf M}_{\text{spin}} = -\frac{eE_{z} \tau \mu_{B}}{\hbar} \sum_{n} \int_{\text{BZ}} \frac{d^{3}k}{(2 \pi)^{3}} \left. \frac{d f}{d \varepsilon} \right|_{\varepsilon = \varepsilon_{n{\bf k}}} \nonumber 
\\
\times \Braket{u_{n{\bf k}} | \frac{\partial H_{\bf k}}{\partial k_{z}} | u_{n{\bf k}}} \Braket{u_{n{\bf k}} | I\otimes{\bf s} |  u_{n{\bf k}}},
\label{spin_magnetization}
\end{eqnarray}
where $\mu_{B}$ is the Bohr magneton and the electron spin $g$-factor $g \approx 2$.
In the present model (\ref{spin_Hamiltonian}), for $\lambda = 0$ the spin magnetization always vanish, 
because of a cancellation between 
the contribution from ($k_{x}$, $k_{y}$, $k_{z}$) and that from ($-k_{x}$, $-k_{y}$, $\pi/c - k_{z}$). This cancellation can be avoided by adding either spin-dependent hoppings within the $xy$ plane (for example, the $\lambda$ term in Eq. (\ref{spin_Hamiltonian})) or spin-independent hoppings between the neighboring layers (for example, the $t_{2}$ term in Eq. (\ref{spinless_Hamiltonian})). 
 }




\section*{Acknowledgments}
This work was supported by Grant-in-Aid for Young Scientists (B) (No. 23740236) , the ``Topological Quantum Phenomena" (No. 25103709) Grant-in-Aid for Scientific Research on Innovative Areas (No.26103006), Grant-in-Aid for Scientific Research (B) (No.26287062), and MEXT Elements Strategy Initiative to Form Core Research Center (TIES) from the Ministry of Education, Culture, Sports, Science and Technology (MEXT) of Japan.


\section*{Author contributions}
T.Y. (T. Yoda) performed numerical calculations based on the tight-binding model.
S.M. conceived and supervised the project.  
T.Y., T.Y. and S.M. discussed the results and wrote the manuscript.
 
\section*{Competing financial interests}
The authors declare no competing financial interests.


\end{document}



\title{Supplementary Note for ``Current-induced Orbital and Spin Magnetization in Crystals with helical structure"}


\author{Taiki Yoda}
\affiliation{Department of Physics, Tokyo Institute of Technology, Tokyo 152-8551, Japan}

\author{Takehito Yokoyama}
\affiliation{Department of Physics, Tokyo Institute of Technology, Tokyo 152-8551, Japan}

\author{Shuichi Murakami}
\email{murakami@stat.phys.titech.ac.jp}
\affiliation{Department of Physics, Tokyo Institute of Technology, Tokyo 152-8551, Japan}
\affiliation{TIES, Tokyo Institute of Technology, 2-12-1 Ookayama, Meguro-ku, Tokyo 152-8551, Japan
}%


\date{\today}

\renewcommand{\thefigure}{S\arabic{figure}}
\renewcommand{\theequation}{S\arabic{equation}}

\maketitle

\section{Properties of the Hamiltonian (1) in {\bf k} space}
\label{sec:aa}
\begin{figure}[h]
\includegraphics[width=6cm]{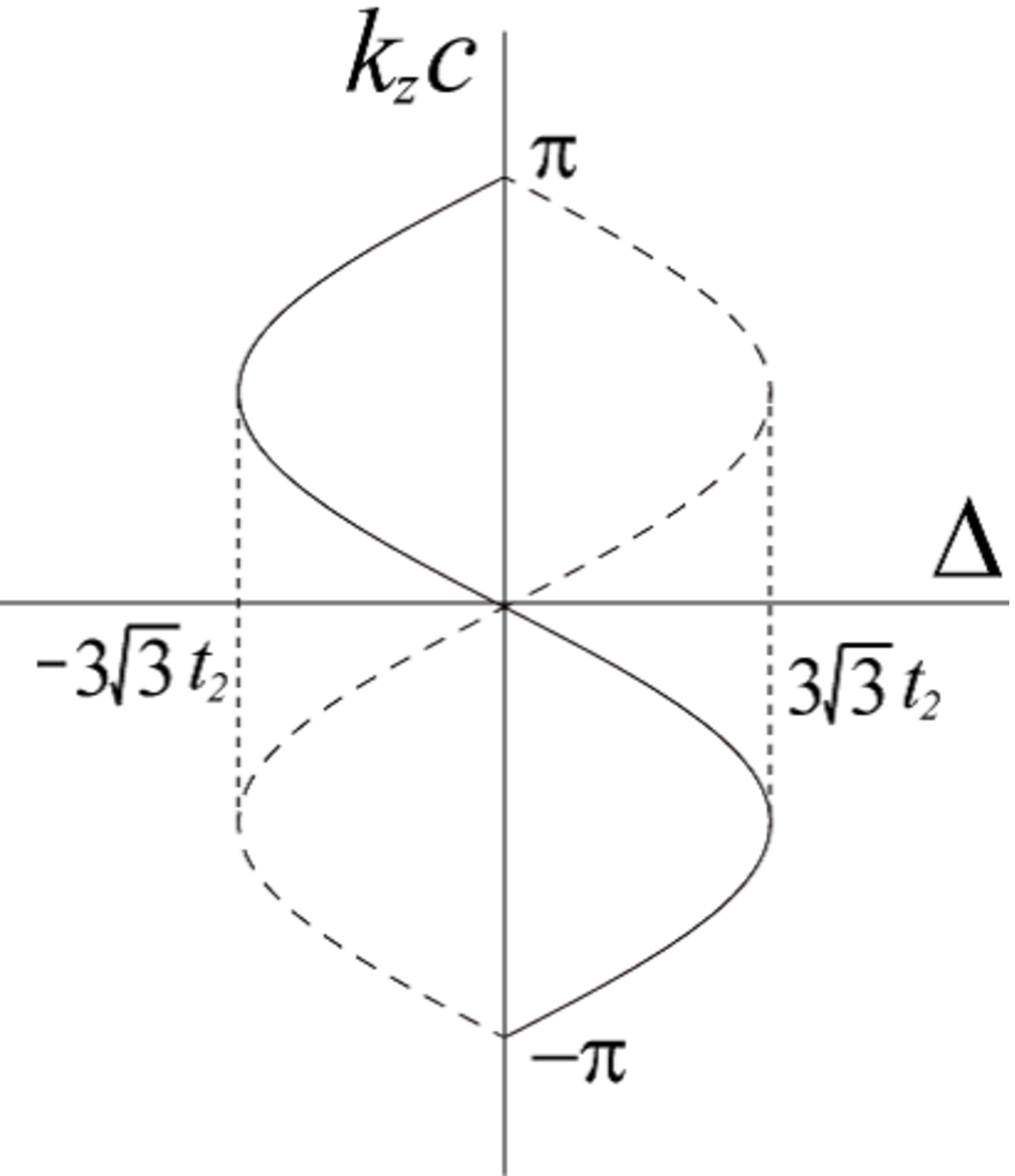}
\caption{{\bf Positions of the Dirac points as a function of $\Delta$}. The Dirac points
(Weyl nodes) appear along the K-H line (solid curve) and the K'-H' line (broken curve).
Only the values of $k_z$ are shown.}
\label{fig:haldane}
\end{figure}

The Hamiltonian (1) for the right-handed helix can be written as
\begin{eqnarray}
H_{{\bf k}} & = & 2t_{2}\cos (k_{z}c) \Bigl[ \sum_{i} \cos ({\bf k}\cdot {\bf b}_{i}) \Bigr] I \nonumber
\\
& + & t_{1} \Bigl[ 1 + \cos({\bf k} \cdot {\bf a}_{1}) + \cos({\bf k} \cdot {\bf a}_{2}) \Bigr] \sigma^{x} \nonumber 
\\
& + & t_{1} \Bigl[ \sin({\bf k} \cdot {\bf a}_{1}) + \sin({\bf k} \cdot {\bf a}_{2}) \Bigr] \sigma^{y} \nonumber 
\\
&+ & \Bigl[\Delta - 2t_{2}\sin (k_{z}c) \sum_{i}\sin({\bf k}\cdot{\bf b}_{i}) \Bigr]\sigma^{z}.
\label{spinless_Hamiltonian_k}
\end{eqnarray}
where $I$ is the $2 \times 2$ identity matrix and $\sigma^{i}$ are the Pauli matrices acting on the sublattice degree of freedom. Vectors ${\bf a}_{1}$ and ${\bf a}_{2}$, defined as ${\bf a}_{1} = a/2(\hat{{\bf x}} + \sqrt3 \hat{{\bf y}})$ and ${\bf a}_{2} = a/2(- \hat{{x}} + \sqrt3 \hat{{y}})$, are the primitive lattice vectors within the $xy$ plane.
As we noted in the main text, the Hamiltonian $H_{{\bf k}}$ becomes the Haldane model on a honeycomb lattice~\cite{PhysRevLett.61.2015} by replacing $k_{z}c$ with a Aharonov-Bohm phase $\phi$ in the second-neighbor hopping in the Haldane model. 
Therefore, the band structure of the Hamiltonian (\ref{spinless_Hamiltonian_k}) is easily obtained from that of the Haldane model~\cite{PhysRevLett.61.2015} by replacing $\phi$ with $k_{z}c$. By analogy with the Haldane model, we assume $|t_{2}/t_{1}| \le 1/3$. For $|\Delta / t_{2}| > 6$, the bands have a finite gap. For $3\sqrt{3} < |\Delta / t_{2}| < 6$, the two bands overlap and are not degenerate; namely systems are in semimetal phases. For $|\Delta / t_{2}| \le 3\sqrt{3}$, the two bands cross each other and form Dirac cones, only on K-H and K'-H' lines in the BZ. On the K-H line, the two bands cross when $\Delta = -3\sqrt3t_{2}\sin k_{z}c$, while on the K'-H' line the two bands cross when $\Delta = 3\sqrt3t_{2}\sin k_{z}c$, as 
shown in Fig.~\ref{fig:haldane}.

\section{Properties of the orbital magnetic moment}
The orbital magnetic moment for the electronic state at ${\bf k}$ of the $n$th band is defined by~\cite{PhysRevB.53.7010,PhysRevB.59.14915}
\begin{equation}
{\bf m}_{n}({\bf k}) = - \frac{ie}{2\hbar} \Braket{ \partial_{{\bf k}} u_{n{\bf k}} | \times [H_{{\bf k}} - \varepsilon_{n{\bf k}}] | \partial_{{\bf k}} u_{n{\bf k}} }.
\end{equation}
The properties of the orbital magnetic moment are similar to those of the Berry curvature~\cite{RevModPhys.82.1959}. If the system has time-reversal symmetry, the orbital magnetic moment satisfies
\begin{equation}
{\bf m}_{n}(-{\bf k}) = - {\bf m}_{n}({\bf k}).
\end{equation}
If the system has spatial inversion symmetry, the orbital magnetic moment satisfies
\begin{equation}
{\bf m}_{n}(-{\bf k}) = {\bf m}_{n}({\bf k}).
\end{equation}
Therefore, unless a system has both time-reversal and inversion symmetries, ${\bf m}_{n}({\bf k})$ is  in general nonzero.

The orbital magnetic moment can be also written as a summation over the eigenstates. The $z$-component of the orbital magnetic moment is calculated as 
\begin{eqnarray}
m_{n,z}({\bf k}) & = & - \frac{ie}{2\hbar} \biggl[ \Braket{ \frac{\partial u_{n{\bf k}}}{\partial k_{x}} | [H_{\bf k} - \varepsilon_{n{\bf k}}]    | \frac{\partial u_{n{\bf k}}}{\partial k_{y}} } - \Braket{ \frac{\partial u_{n{\bf k}}}{\partial k_{y}} | [H_{\bf k} - \varepsilon_{n{\bf k}}]    | \frac{\partial u_{n{\bf k}}}{\partial k_{x}} } \biggr] \nonumber 
\\
& = & \frac{e}{\hbar} \text{Im} \Braket{ \frac{\partial u_{n{\bf k}}}{\partial k_{x}} | [H_{\bf k} - \varepsilon_{n{\bf k}}] | \frac{\partial u_{n{\bf k}}}{\partial k_{y}} } \nonumber
\\
& = & \frac{e}{\hbar}\text{Im} \sum_{m \neq n} \frac{ \Braket{ u_{n{\bf k}} | \frac{ \partial H_{{\bf k}} }{ \partial k_{x} } | u_{m{\bf k}} } \Braket{ u_{m{\bf k}} | \frac{ \partial H_{\bf k} }{ \partial k_{y} } | u_{n{\bf k}} } }{ (\varepsilon_{m{\bf k}} - \varepsilon_{n{\bf k}}) }.
\label{orbital_magnetic_moment_z}
\end{eqnarray}
The final expression does not include derivatives of the eigenstates, which is useful for numerical calculations. Equation (\ref{orbital_magnetic_moment_z}) shows that ${\bf m}_{n}({\bf k})$ diverges when $\varepsilon_{n{\bf k}} = \varepsilon_{m{\bf k}}$ at certain values of ${\bf k}$. Therefore, ${\bf m}_{n}({\bf k})$ is enhanced around degeneracy points. 

\begin{figure}
\includegraphics[width=16cm]{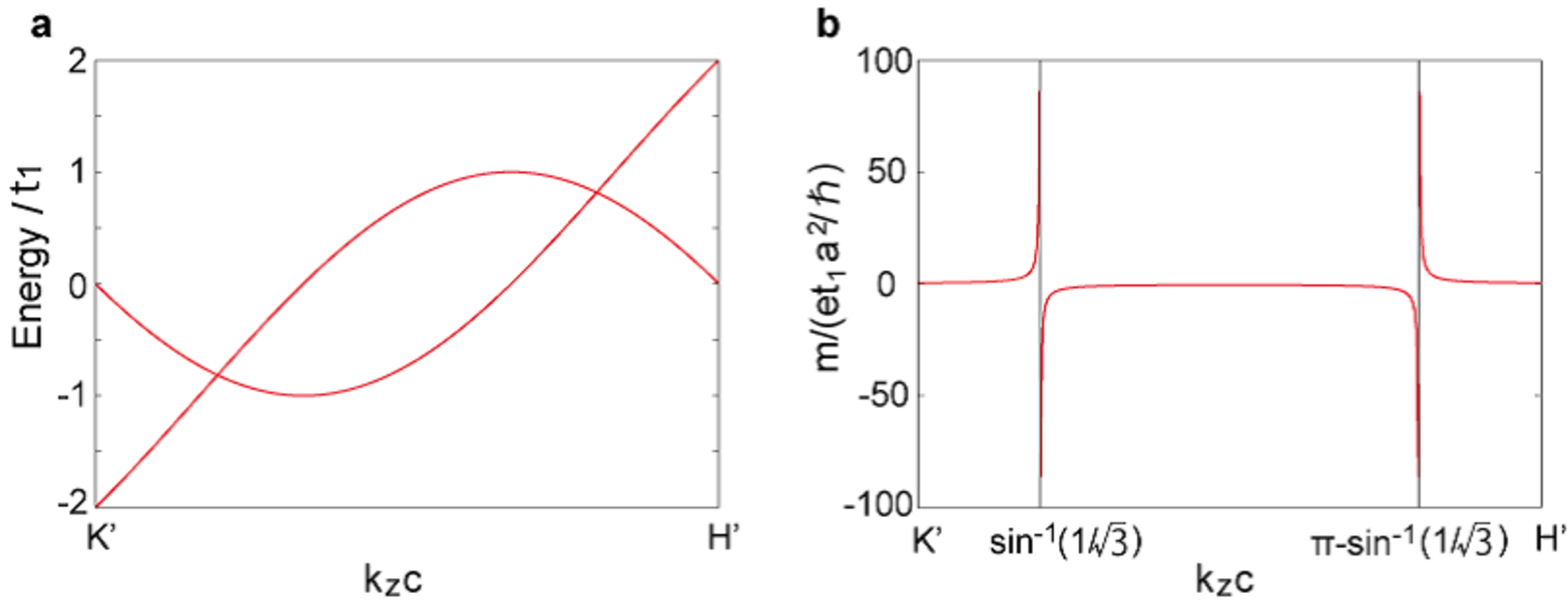}%
\caption{{\bf a}, Energy band for the Hamiltonian (\ref{spinless_Hamiltonian_k}) along the K'-H' lines. The parameters are $t_{2} = t_1/3$, and $\Delta = t_1$. {\bf b}, The z-component of the orbital magnetic moment on the K'-H' line.}
\label{Energy_band_and_magnetic_moment}
\end{figure}

We now consider the Hamiltonian (\ref{spinless_Hamiltonian_k}). The orbital magnetic moments of the two bands are identical as seen from Eq.(\ref{orbital_magnetic_moment_z}). Figure \ref{Energy_band_and_magnetic_moment}{\bf a} shows the energy band of the Hamiltonian (\ref{spinless_Hamiltonian_k}) on the K'-H' line and Fig. \ref{Energy_band_and_magnetic_moment}{\bf b} shows the $z$-component of the orbital magnetic moment on the K'-H' line. Dirac points appear at $k_{z}c = \sin^{-1}(1/\sqrt3)$ and $\pi - \sin^{-1}(1/\sqrt3)$. We can see that the orbital magnetic moment is enhanced around the Dirac points and diverges at the Dirac points. The enhanced orbital magnetic moment around the Dirac points makes a large contribution to the orbital magnetization.

\section{{\bf k}-space representation of the Hamiltonian (4)}
The Hamiltonian $H_{\text{so}}$ for the right-handed helix can be written in {\bf k}-space as
\begin{eqnarray}
	H_{\text{so},{\bf k}} = \begin{pmatrix}
		H_{\text{A}} & H_{\text{AB}} \\
		H_{\text{AB}}^{\dagger} & H_{\text{B}}
					\end{pmatrix}.
\end{eqnarray}
The $2 \times 2$ matrices $H_{\text{A}}$, $H_{\text{B}}$, and $H_{\text{AB}}$ are
\begin{eqnarray}
H_{\text{A}} & = & \Delta I + \lambda_{xy} [ 2\sin ( {\bf k} \cdot {\bf b}_{1} + k_{z}c ) - \sin ( {\bf k} \cdot {\bf b}_{2} + k_{z}c ) - 					\sin({\bf k} \cdot {\bf b}_{3} + k_{z}c ) ] s^{x} \nonumber
\\ 
& + & \sqrt3 \lambda_{xy} [ \sin( {\bf k} \cdot {\bf b}_{2} + k_{z}c ) - \sin( {\bf k} \cdot {\bf b}_{3} + k_{z}c) ] s^{y} \nonumber
\\
& + & 2\lambda_{z} [ \sum_{i} \sin( {\bf k} \cdot {\bf b}_{i} + k_{z}c ) ] s^{z},
\\
H_{\text{B}} & = & - \Delta I -\lambda_{xy} [ 2\sin( -{\bf k} \cdot {\bf b}_{1} + k_{z}c ) - \sin( -{\bf k} \cdot {\bf b}_{2} + k_{z}c ) - \sin( -{\bf k} \cdot {\bf b}_{3} + k_{z}c ) ] s^{x} \nonumber
\\
& - & \sqrt3 \lambda_{xy} [ \sin( -{\bf k } \cdot {\bf b}_{2} + k_{z}c ) - \sin( -{\bf k} \cdot {\bf b}_{3} + k_{z}c ) ] s^{y} \nonumber
\\
& + & 2\lambda_{z} [ \sum_{i} \sin( -{\bf k} \cdot {\bf b}_{i} + k_{z}c ) ] s^{z},
\\
H_{\text{AB}} & = & t_{1} [ 1 + e^{ -i {\bf k} \cdot {\bf a}_{1} } + e^{ -i {\bf k} \cdot {\bf a}_{2} } ] I \nonumber
\\
& + & \frac{\sqrt3}{2} i\lambda [ e^{ -i {\bf k} \cdot {\bf a}_{1} } - e^{ -i{\bf k} \cdot {\bf a}_{2} } ] s^{x}
- \frac{1}{2} i\lambda [ 2 - e^{ -i {\bf k} \cdot {\bf a}_{1} } - e^{ -i {\bf k} \cdot {\bf a}_{2} } ] s^{y},
\end{eqnarray}
where $I$ is the $2 \times 2$ identity matrix and $s^{i}$ are the Pauli matrices acting on spin space.